# The Multiply Scattering Effect on the Energy Measurement of UHE Cosmic Rays using Atmospheric Fluorescence Technique

**Xingzhi Zhang***

*Department of Physics, Columbia University, New York, NY 10027,USA*

Feb. 29 2000

Abstract

Point sources in the atmosphere are surrounded by aureole because of atmospheric scattering. The properties of the time-dependent aureole radiance are calculated by use of a Monte Carlo approach and an iterative method. Since the aureole is particularly important in the ultraviolet, which is the region the Ultra-High-Energy (UHE) cosmic ray experiment using the air fluorescence technique like Fly's Eye or High-Resolution-Fly's-Eye(HiRes) are set in. The effect of the multiply scatteing on the energy measurement is studied.

∗*currently visiting Department of Physics, University of Utah, SLC UT84112*

*email: xz50@columbia.edu*

*URL: http://www.columbia.edu/~ xz50*

Fax: 1-801-581-6256

Phone: 1-801-581-4306

# 1 Introduction

Every well determined feature of the cosmic ray energy spectrum will have considerable impact on theories of the origin, acceleration and propagation of cosmic rays. After 40 years effort by many groups, the details of the energy spectrum above $10^{17}$ eV are still limited by statistics, systematics and resolution. Experimentalists have been searching for the existence of a cut off on the energy spectrum above $10^{20}$ eV for more than 30 years. This cutoff could result from the interaction of cosmic ray protons or nuclei and the $2.7^o$ K black body radiation [1] if the sources are distant enough.

The detection of these extremely high energy cosmic rays is necessarily indirect because of the extremely low flux. The earth's atmosphere makes the low flux detectable by converting the cosmic ray primaries into extensive air showers(EAS) of various secondary particles at ground level by detecting the secondary particles or the Cerenkov light they produce. Alternatively, one can detect the atmospheric nitrogen fluorescence light induced by the passage of the shower. This technique employed by the Fly's Eye detector [2] and its successor the High Resolution Fly's Eye (HiRes) [3] are the only way capable of measuring longitudinal shower developments individually, thus allowing a direct estimation of each shower's primary energy.

The basic mechanism contribute to the generation of the light signal seen by Fly's Eye or HiRes detector is nitrogen fluorescence light which relates directly to the number of charged particles in an EAS. The fluorescence light is emitted isotropically from the shower, allowing for detection of showers at large distances. In HiRes detector, UHE shower can bee seen as far as 40 km's away. This makes a very big aperture, and has high statistics for the cosmic ray energy spectrum. The disadvantage is that, the multiply scattering effect will become important when the shower is far away. In this paper, we will study this effect on the energy estimation of this technique.

The multiply scattering of light by molecules and aerosols in the atmosphere gives rise to a radiance field about a pointlike source. This contribution, or called aureole, depends on many parameters, such as the single scattering albedo, the optical thickness, the scattering phase function, the off angle of the detector from the source direction, the field view solid angle of the detector, and also the integration time window of the detector. This effect has been studied for the last 25 years by the people who are interested in solar blind ultraviolet communication and warning systems or atmospheric study. However most studies were devoted mainly to the total aureole effect [4] [5] [6]. There were few studies on the time-dependent aureole effect. But they were either for the case of very large optical thickness (fog, cloud, etc) [7] or only for very low order of scattering which is for the case of small optical thickness [8] [9].

Since the shower detected by HiRes can be in the range from a few km to more than 40 km, on good weather ( less aerosol) or bad weather ( dense aerosol). And also the detector is integrating the signal within about 5 micro second time window. We will present calculations of the time-dependent aureole radiance field about an impulsive isotropic source in a scattering and absorbing medium. The approach used is generally applicable to any case, and does not involve any approximation. The calculation of the temporal characteristics of the scattering radiation are based on the work of Trakhovsky *et al.* [9].

In Section 2 we will present the recursive approach to calculate the scattering radiation effect initially developed by Zachor [4], and generalized by Trakhovsky *et al* [9] to time-dependent case. In Section 3 we will present the Monte Carlo way to calculate the temporal scattering radiation. The results are given in Section 4 for both cases in different atmosphere conditions and detector setup.

And then we apply the result for any cosmic ray shower detected by HiRes detector, and calculate the effect on the energy estimate. This is done in Section 5.

## 2 Recursive Approach

As shown in Fig. 1. The first-order scattering term can be decomposed into three steps: direct transmission from the source to the volume element $dV$, scattering inside $dV$ in the detector direction, and direct transmission from $dV$ to the detector. Assume an isotropic point source which emits at an instant t = 0 an inpuls of total photo Q. The irradiance (power/unit area) incident on element $dV = r_2^2 dr_2 d\Omega$ is:

$$\mathcal{E} = \frac{Q}{4\pi r_1^2} exp[-\alpha + \beta)r_1]. \tag{1}$$

in which $\beta$ is the volume scattering coefficient, $\alpha$ is the absorption coefficient. Accordind to the definition of scattering coefficient $\beta$ and single-scattering phase function $P\left(cos\theta\right)$, the scattered intensity of radiation by $dV$ toward detector is:

$$d\mathcal{I}_s = \mathcal{E}\beta P(cos\theta)dV. \tag{2}$$

The irradiance incident on the detector is:

$$d\mathcal{E}_s = \frac{d\mathcal{I}_s}{r_2^2} exp[-(\alpha + \beta)r_2]. \tag{3}$$

Then the radiance (photo/unit area/unit solid angle) received at O is:

$$d\mathcal{N}_s = \frac{d\mathcal{E}_s}{d\Omega} = \frac{Q\beta P\left(cos\theta\right)}{4\pi {r_1}^2} exp\left[-\left(\alpha + \beta\right)\left(r_1 + r_2\right)\right] dr_2 \tag{4}$$

Since we are interested in a time-resolved measurement. Following Reilly and Warde [8]. we chose a prolate spheroidal coordinate system with a source positioned at one local point and a detector at the other. It may be shown [8] that the parameters of Fig. 1. $(r_1, r_2, \theta, \gamma)$ are transformed into prolate spheroidal coordinates using the following relationships:

$$r_1 = \frac{R}{2}(\xi + \eta), \tag{5}$$

$$r_2 = \frac{R}{2}(\xi - \eta), \tag{6}$$

$$cos\theta = \frac{2 - \xi^2 - \eta^2}{\xi^2 - \eta^2}, \tag{7}$$

$$\gamma = cos^{-1}\left(\frac{1 - \xi\eta}{\xi - \eta}\right). \tag{8}$$

let $ct = r_1 + r_2$. $t$ is the scattered photon time of flight. Using Eqs. (5) and (6) we obtain:

$$\xi = \frac{r_1 + r_2}{R} = \frac{ct}{R}, \tag{9}$$

Transforming Eq. (8) we have

$$\eta = \frac{1 - \xi cos\gamma}{\xi - cos\gamma}, \tag{10}$$

Using above Eqs. then Eq. (4) can be mathematically transferred to

$$d\mathcal{N}_s(R, \gamma, t) = \frac{cQ\beta P\{cos[\theta(R, \gamma, t)]\}}{2\pi R^2} \bullet \frac{exp[-(\alpha + \beta)ct]}{\xi^2 - 2\xi cos\gamma + 1} dt. \tag{11}$$

Where $\theta$ is calculated by Eqs. (7), (9) and (10). Let

$$N_1(R, \gamma, t) = \frac{d\mathcal{N}_s(R, \gamma, t)}{dt}. \tag{12}$$

is defined as the temporal radiance. In order to remove the singularity in Eq. (11), and also for conveniency of calculating high order scattering, we define the apparent temporal radiance as:

$$B_n(R,\gamma,t) = \frac{4\pi R^2 sin\gamma}{Q}\sigma^{-n}N_n(R,\gamma,t) \tag{13}$$

Then Eq. (11) can be transferred to:

$$B_1(R,\gamma,t) = ck_{ext.}\frac{2P\{cos[\theta(R,\gamma,t)]\}sin\gamma}{(\xi^2 - 2\xi cos\gamma + 1)}exp(-k_{ext.}ct). \tag{14}$$

where $k_{ext.} = \alpha + \beta$ is the total extinction coefficiency, $\sigma = \frac{\beta}{\alpha+\beta} = \frac{\beta}{k_{ext.}}$ is the single-scattering albedo.

Following the same way as [9], assume the apparent temporal radiance of *(n-1)* th order incidented at instant $t'$ at an angle $\gamma'$ on a volume element $dV'$ located at distance $R'$ from the source through a field view solid angle $d\Omega'(\gamma',\phi')$ (see Fig. 2) is $B_{n-1}(R',\gamma',t')$.

Geometrically it is easy to get the following relationship:

$$R' = (R^2 + r'^2 - 2Rr'cos\gamma)^{1/2}. \tag{15}$$

$$\epsilon' = cot^{-1}\left(\frac{cos\gamma - \frac{r'}{R}}{sin\gamma}\right), \tag{16}$$

The scattering angle $\theta'$ is defined [4] by

$$cos\theta' = cos\gamma' cos\epsilon' + sin\gamma' sin\epsilon' cos\phi'. \tag{17}$$

then the $n$th order apparent temporal radiance is

$$B_n(R,\gamma,t) = sin\gamma\int_0^{D'} dr'\int_0^{\pi} d\gamma'\left(\frac{R}{R'}\right)^2 B_{n-1}(R',\gamma',t')\overline{P(cos\theta')}. \tag{18}$$

where $\overline{P(cos\theta')}$ is an azimuthally integrated scattering phase function [4]:

$$\overline{P(cos\theta')} = \int_0^{2\pi} d\phi' P(cos\theta'). \tag{19}$$

where

$$D' = \frac{R}{2}(\xi - \eta). \tag{20}$$

$$SB + BC = ct' = ct - r'. \tag{21}$$

$\xi, \eta$ is defined in Eqs. (9) and (10).

With the above equations, theoretically we can calculate any order of scattering radiance for any given geometry setup (R,$\gamma$,t) and any atmosphere ( $\sigma$, $P(cos\theta)$, $k_{ext.}$ ). But in the real calculation, it will take tremendous of CPU time to calculate the radiance high than 3rd order of scattering, because of the 3-dimension integration in Eq. (18).

The total apparent temporal scattering radiance is:

$$B(R, \gamma, t) = \sum_{n=1}^{n=\infty} \sigma^n B_n(R, \gamma, t) \tag{22}$$

For simplification purpose, we define

$$c = k_{ext.} = 1. \tag{23}$$

Then in the following, both the distance and time are in the unit of extinction length.

We then define regular grids for both the geometry setup and the atmosphere: R is from 0.0 to 6.0 with interval of 0.1; $\gamma$ is from $0.01^o$ to $179.99^o$ with interval $0.2^o$ when $\gamma \leq 2^o$, with interval $2^o$ when $\gamma > 2^o$; $t - R$ is from 0 to 2 with interval 0.1; $\sigma$ is from 0.0 to 1.0 with interval 0.2; The atmospheric phase function is a weighted average of single-scattering phase functions $P_R(cos\theta)$ and $P_A(cos\theta)$, which represent, respectively, the Rayleigh and aerosol components. The weights are the corresponding scattering coefficients:

$$P(cos(\theta)) = \frac{\beta_R P_R(cos(\theta)) + \beta_A P_A(cos(\theta))}{\beta_R + \beta_A} = \rho P_R(cos(\theta)) + (1-\rho) P_A(cos(\theta)). \quad (24)$$

$$\rho = \frac{\beta_R}{\beta_R + \beta_A}. \quad (25)$$

$\rho$ is from 0.0 to 1.0 with interval of 0.2.

The grids chosed above to calculate the scattering radiance are simply trying to cover all the different situations the HiRes experiment will meet, and also considering the CPU time and calculation errors. To calculate the *n*th order, we only need to interpolate the result of the *(n-1)*th order, and then calculate the integration in Eq. (18) by using Gauss-Legendre procedure. The error of this calculation is about a few percent.

## 3 Monte Carlo Calculation

We made Monte Carlo calculations for the scattering radiance also. There are many references on this method [6]. Here we follow the way as [6]. The basic idea of MC method is to decompose light into a set of pencils of light that are called photons for the sake of brevity. The program follows the path of each photon inside the medium. Since the source is an isotropic point source, so it is very simple to setup the MC process (it can be generated to more completed case):

The geometry for the MC is shown in Fig. 3. Source S is assumed to be placed at the center of a sphere of radius $R_{max}$. The photon trajectory is then simulated by successive straight lines between collisions with scattering and absorbing centers inside the sphere of radius $R_{max}$. Each interaction between a photon and a scattering center obeys the law of single scattering. The scattering coefficient is used in computing the probability of the photon's being scattered between distance $l$ to $(l+dl)$ as $\beta exp(-\beta l)dl$. To take into account of absorption along the path, we weight a photon by factor $W_i$ that is initially set at 1 and then multiplied by $exp(-\alpha l)$ at each collision. After the scattering, the new direction of the photon is determined by the scattering phase function $P(cos\theta)$. To take the advantage of the spherical symmetry, we look the whole spherical surface with radius $R$ as a detector with normal direction.

When a photon crosses the detector's surface, we record 2 quantities: the real path length $ct$, the angle $\gamma$ between the photon's direction and the local normal direction of the detector. Then the weight of the photon $W_i$ is:

$$W_i = exp(-\alpha ct). \tag{26}$$

If a photon does not hit the detector at time $t \longrightarrow t + dt$, within direction $\gamma \longrightarrow \gamma + d\gamma$, then

$$W_i = 0. \tag{27}$$

The total number of photons hit the detector at time $t \longrightarrow t + dt$, within direction $\gamma \longrightarrow \gamma + d\gamma$, with the detector aerie $dS$ is:

$$\mathcal{B}_{mc}(R, \gamma, t)dtd\Omega dS = \sum_{i=1}^{i=Q_T} W_i(R, \gamma, t) \tag{28}$$

where $Q_T$ is the total number of input photons. (in the real calculation we simulate up to $10^8$ photons). From the spherical symmetry of the detector, it is easy to get the detector aerie:

$$dS = 4\pi R^2 cos(\gamma). \tag{29}$$

and the field view solid angle $d\Omega$:

$$d\Omega = 2\pi \int_{\gamma}^{\gamma+d\gamma} sin(\gamma')d\gamma'. \tag{30}$$

Same as the definition in Eqs. (13) and (22) From Eqs. (26), (27), (28), (29), (30) we get the apparent temporal radiance:

$$B_{mc}(R,\gamma,t) = \frac{4\pi R^2 sin\gamma}{Q_T}\frac{\mathcal{B}_{mc}(R,\gamma,t)dtd\Omega dS}{dtd\Omega dS} = \frac{\sum_{i=1}^{i=Q_T} W_i(R,\gamma,t)}{Q_T}\frac{sin\gamma}{cos\gamma}\frac{1}{2\pi\int_{\gamma}^{\gamma+d\gamma} sin(\gamma')d\gamma'}. \quad (31)$$

In order to get good statistics, we simulate as many photons as possible within reasonable CPU time.

As described in Section 1. the atmosphere is completely defined by extinction coefficient $k_{ext}$, single-scattering albedo $\sigma$, and total phase function $P(cos\theta)$ which is described in Eqs. (24), (25). In the following we will describe how we model the phase functions.

The Rayleigh scattering phase function is modeled as the simplified form:

$$P_R(cos\theta) = \frac{3}{16\pi}(1+cos^2\theta) \quad (32)$$

Usually the aerosol phase function is described by a modified Henyey-Greenstein function, with an additional parameter $f$ that gives rise to a backward peak:

$$P_A(cos\theta) = \frac{1-g^2}{4\pi}[\frac{1}{(1+g^2-2gcos\theta)^{3/2}} + \frac{f(3cos^2\theta-1)}{2(1+g^2)^{3/2}}]. \quad (33)$$

where $g$ is asymmetry parameter.

Since we are interested in the HiRes experiment, which is set in the desert of western USA. Here we will use a desert aerosol phase function calculated from Mie scattering theory with a aerosol particle size distribution function $a^{-4}$, where $a$ is aerosol particle size. The phase functions are shown in Fig. 4. This phase function is very close to the real aerosol phase function at HiRes site [10].

## 4 Results

The recursive calculation was performed for up to 15 order for 6 optical depth.

In Figs. 5(a), 5(b), we show the first a few order of scattering for $\sigma = 0.8$, $\rho = 0.8$ $R = 1$, and *time* is defined as $t - R$, which is the traveling time of scattering light $t$ different from the direct light traveling time $R$. In 5(a) $\gamma = 2^o$, in 5(b) $\gamma = 20^o$. The signal is defined from Eqs. (18), (22), and the unit has been scaled by $1/4\pi$. In Figs. 6(a), 6(b), we show the first a few order of scattering for the same condition as Figs. 5. except $R = 4$. From these Figs. we can see when the view angle $\gamma$ become bigger, or source detector distance $R$ becomes longer, the high order scattering becomes important.

In Figs. 7(a), 7(b), we show the dependency of the total scattering radiance on the single scattering albedo $\sigma$. $\sigma$ is from 0.2 to 1.0 from bottom to up (when $\sigma$ is 0.0 there is not scattering, so we did not plot it out). The other conditions are $\rho = 0.8$, $R = 1$, 7(a) $\gamma = 2^o$, in 7(b) $\gamma = 20^o$. It is very clear when the single scattering albedo is bigger the scattering is stronger. In Figs. 8(a), 8(b), we show the dependency of the total scattering radiance on the rate of aerosol scattering coefficiency $\rho$. $\rho$ is from 0.0 to 1.0 from top to bottom. $R = 1$, $\sigma = 0.8$, 8(a) $\gamma = 2^o$, 8(b) $\gamma = 20^o$. Remember the aerosol phase function is more forward distributed( see Fig. 4). This can explain why the scattering is stranger when $\rho$ is high at small time range.

In Figs. 9(a), 9(b), we show the comparisons of the result from Monte Carlo to the iterative method. for $\sigma = 0.8$, $\rho = 0.8$ $R = 1$, In 9(a) $\gamma = 3^o$, in 9(b) $\gamma = 19^o$. In Figs. 10(a), 10(b), we show the same thing as Fig 9. except $R = 4$. The difference between the two methods are a few percent. except at large optical depth, where it reachs 10%. The difference mainly come from the iterative calculation, as pointed out in Sec. 2, by the interplating and Gauss-Legendre integration. The statistical fluctuations of MC results is poor even we used 48 hours of CPU on an 500-MHZ machine. From application point view, the iterative calculation is a more useful way. The error from iterative calculation can be lowed by more notes of interplation and integration. As will be pointed out later in Set. 5. the current result is good enough for the application on HiRes experiment.

# 5 Application

In this section, we will use the above iterative calculation results to study the multiply scattering effect in the energy estimate of cosmic ray by fluorescence technique, such as HiRes experiment. In the past 10 years, HiRes group has developed a detailed MC code to simulate the detector response for any given EAS.

The EAS was generated from CORSIKA package [11] [12]. The basical idea for the HiRes detector MC is that, start from the input shower, calculate the light produced at every stage, which includes fluorescence light and Cerenkov light. The fluorescence light then transmit to the detector. The Cerenkov light is mainly in the direction of the input shower, which can be scattered into the detector by Rayleigh scattering or aerosol scattering. The fluorescence light is directly related to number of charged particle at given stage, but the Cerenkov light is more complicated, and related to the history of the shower development. Fortunately, the Cerenkov light is much weak than the fluorescence light when the shower does not directly shoot to the detector. When the shower are within the direction of the detector, we simply drop it for better understanding of the data. All the good events we chose is only fluorescence light dominated. Then the above calculation can be directly used here.

Let $\mathcal{Q}(x)dx$ is the total fluorescence light generated at stage $x \longrightarrow x + dx$, *F(x)* is the detector coefficiency relate to the source at point *x*. *r(x)* is the distance from the source to the detector.

Then the direct light from stage $x \longrightarrow x + dx$ to the detector is:

$$S_{dir.}(x) = F(x) \frac{\mathcal{Q}(x)dx}{4\pi r^2(x)} exp(-k_{ext.} r(x)). \tag{34}$$

If we define the shower start at time $t = 0$, the time the direct light flight to the detector is: (Fig. 11.)

$$t(x) = \frac{r(x) + l(x)}{c} \tag{35}$$

let

$$c = k_{ext.} = 1 \tag{36}$$

Since the electronics of the detector will integrate all the signal come to the detector from point $x$ within field view solid angle $\Omega(x)$, within a fixed time window $W$, the scattering light from point $x'$ will also be integrated if it is in the time window and in the right direction. The geometric parameters for calculation of this effect is in Fig. 11. This effect can be represented as:

$$S_{sct.}(x) = F(x) \int_{t(x') \leq t(x)+W/2} dx' \int_{t(x)-W/2}^{t(x)+W/2} dt' \int_{0}^{\Omega(x)} d\Omega \frac{\mathcal{Q}(x')}{4\pi r^2(x') sin(\gamma(x', x))} B[r(x'), \gamma(x', x), t']. \tag{37}$$

in which $B(r(x'), \gamma(x', x), t)$ is calculated from Eqs. (14), (18), (22).

In the real calculation, we first build up a table about $B(r(x'), \gamma(x', x), t)$ as described in Sec. 2. For Eq. (37), we simple interplate the result from the table for any given geometry $r(x'), \gamma(x', x), t$ at any atmosphere ( $k_{ext.}, \sigma$ ) condition. We chose $W$ as 5.6 micro second, which is the real time integration window for HiRes detector. The detector openning field view solid angle $\Omega(x)$ is a lit complicated, which is related to the stage of the shower and also the detector pixel.

The Calculation also depends on the atmosphere model, the following results based on the standard US atmosphere model. The aerosol part based on an exponentially decay model with an scale height of 1.2 Km, the extinction length on the ground is 10 Km.

The results are show in Figs. 12, 13, 14 15. for vertical EAS showers with primary energy $10^{20}$ eV at 10 km, 20 km, 30 km and 40 km away from the detector respectively. In 12(a), 13(a), 14(a), 15(a) we show the longitudinal shower profile at every stage from top of atmosphere to the ground. The solid lines are the total signal, the dashed lines are the direct light, the dot lines show the multiply scattering light. The X axis is the zenith angle for given stage, the Y axis is the signal measured at the detector site. Because of the detector trigger threshold condition, only the parts of shower around $\pm 10^o$ from the maximum signal point will be seen by the detector and gotten trigged.

In 12(b), 13(b), 14(b), 15(b) we show the rate of the multiply scattering light contribution to the directly light, define as $S_{sct.}(x)$ / $S_{dir.}(x)$ from Eqs. (34), (37). We can see the effect is around 10% within the range the detector can see, and becomes stronger when the shower is farther. At the stage close to the ground, because both the atmosphere and aerosol density is higher, the extinction length is shorter, so the optical depth is longer the multiply scattering effect becomes stronger. For 12(b), the multiply scattering effect at early stage is stronger, this is simply because the geometrical distance is longer for early stage than later stage with a factor $1/sin\theta$ (Fig. 11.), which makes the optical depth longer even when the density of atmosphere become lower when it go up. The thickness of atmosphere is fixed, when the shower is farther like 13(b), 14(b) and 15(b). this geometrical effect becomes less important.

In order to see directly how the multiply scattering effect contribute to the energy estimate of this experiment. We draw the longitudinal profile versus the shower depth X in unit of $gm/cm^2$, as Fig (16). The shower to detector distance is 30 Km away. The aeries under the lines are directly proportional to the primary energy of the Cosmic ray [12] as:

$$E_{primary} \propto \int dX sig(X). \tag{38}$$

We can easily get the contribution to the primary energy estimation. The result shown in Fig. (17) is the rate of the aerie from scattering light to the aerie from the total light in percentage. Here we also studied how much this effect is, based on different aerosol models. We changed the horizontal aerosol extinction length from 10 Km to 20 Km, with the scale height change from 1.2 Km to 1.5 Km. We can see the scatteing light contributes to the energy estimation in about 10%.

It is interesting to see that when the shower to detector distance is moved from 10 Km to 40 Km, the scattering effect does not increase dramatically. This is because we only intergrate the signal within a fixed time window.

More generally, the EAS can not be verticle, we also simulated showers with arbitrary zenith angle, and the multiply scattering effect on the shower longitudinal profile will change shower by shower depend on the shower gemetry. But the effect on the primary energy estimate is still about 10%. In the real HiRes data analysis, we will calculate this effect for every shower.

As point in Sec. 4, there is about 10% error in the scattering radiance calculation. But since the multiply scattering light itself is only about 10% for the total light, so the error transferred from the scaterring radiance calculation is only about 1%, which is much small than the systematics of this experiment.

## 6 conclusion

The properties of an aureole about a point source that are due to atmospheric scattering were calculated by two approaches: the Monte Carlo method and iterative technique. For the iterative technique we have an applicable approach to calculate as high order of scattering as you want within reasonable calculation time for any kind of atmosphere and any detector setup. The results match very well with the MC results within error of about 10%. The iterative method is more useful than MC method if we are interested in the temporal aureole problem. We studied the properties of the temporal aureole depend mainly on single-scattering albedo, scattering phase function, optical depth between source and detector, detector view angle.

Then we apply the multiply scattering property of a point source to the EHE cosmic ray shower detected by the atmosphere fluorescence technique. And found the contribution of the scattering light is about 10% , which depends on how far the shower is from the detector in the unit of optical depth. This means the energy estimate will be corrected by 10% because this effect.

This approach can also be generalized to anisotropic cases. Such as the Lidar system used to calibrate the atmosphere [13]. In the world there are some Cerenkov light detector like [14]. Since for anisotropic case the multiply scattering effect will become stronger [6], in order to understand better of the system, the multiply scattering effect should be studied in those cases.

## 7 Acknowledgments

The author thanks Prof. W. Lee, Prof. P. Sokolsky, Dr. c. Zen, Dr. L. Wiencke for the creative discussions. The support from High Energy Astrophysics Institute at University of Utah is gratefully acknowledged during his visiting time.

## References

[1] K. Greisen, Phys. Rev. Lett. **16**, 748 (1966); G. T. Zatsepin and V. A. Kuzmin, Pis'ma Zh. Eksp. Teor. Fiz. **4**. 114 (1966) [JETP Lett. **4**. 78 (1966)].

[2] R. M. Baltrusaitis *et al.*, Nucl. Instrum. Methods Phys. Res., Sect. **A 240**, 410 (1985); G. L. Cassiday, Annu. Rev. Nucl. Part. Sci. **35**, 321 (1985).

[3] T. Abu-Zayyad *et al*, to be published in Phys. Rev. Lett.

[4] A. S. Zachor, Appl. Opt. **17**, 1911 (1978).

[5] F. Riewe and A. E. S. Green, Appl. Opt. **17**, 1923(1978).

[6] C. Lavigne *et al.*, Appl. Opt. **38**, 6237(1999). and references within.

[7] L. B. Stotts, Appl. Opt. **17**, 504(1978); K. Furutsu, J. Opt. Soc. Am. **70**, 360 (1980); G. C. Mooradian and M. Geller, Appl. Opt. **21**, 1572(1982).

[8] D. M. Reilly and C. Warde, J. Opt. Soc. Am. **69**, 464(1979).

[9] E. Trakhovsky and U. P. Oppenheim, Appl. Opt. **22**, 1633 (1983).

[10] XZ. Zhang, Ph.D thesis, Columbia University (2000) unpublished.

[11] D. Heck *et al.*, Report FZKA 6019 (1998), Forschungszentrum Karlsruhe.

[12] C. Song *et al.* (submitted to Astr. Part. Phys. 2000).

[13] T. Abu-Zayyad *et al*, ICRC 1999(Salt Lake City 1999), OG4.5.10.

[14] Urban, M., et al. 1996, NIM A **368**, 503

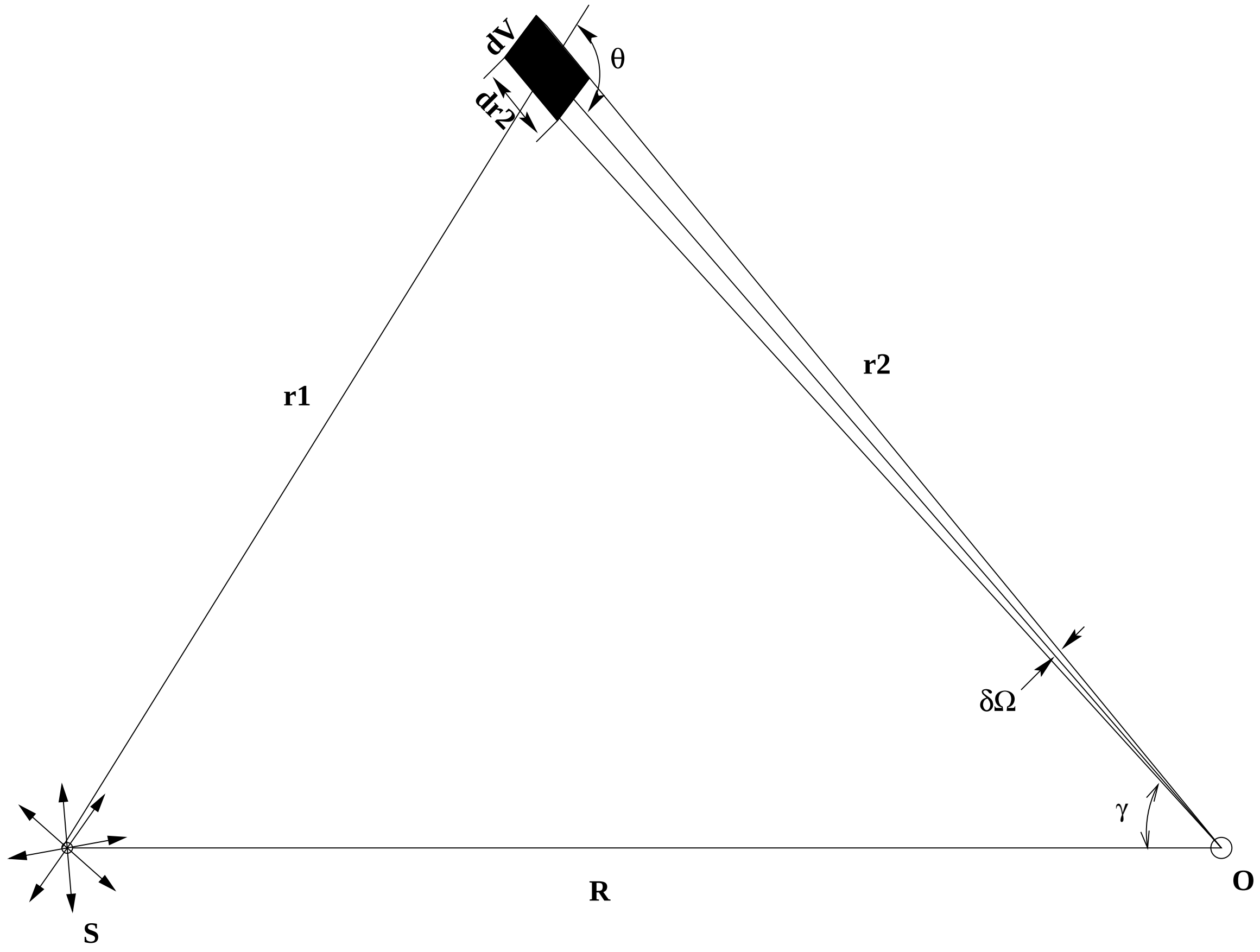

Figure 1: Geometric parameters involved in the calculations of first order scattering. S is the source point; O is the detector point. $\gamma$ is the off angle; $\theta$ is the scattering angle; $\delta\Omega$ is the detector field view solid angle

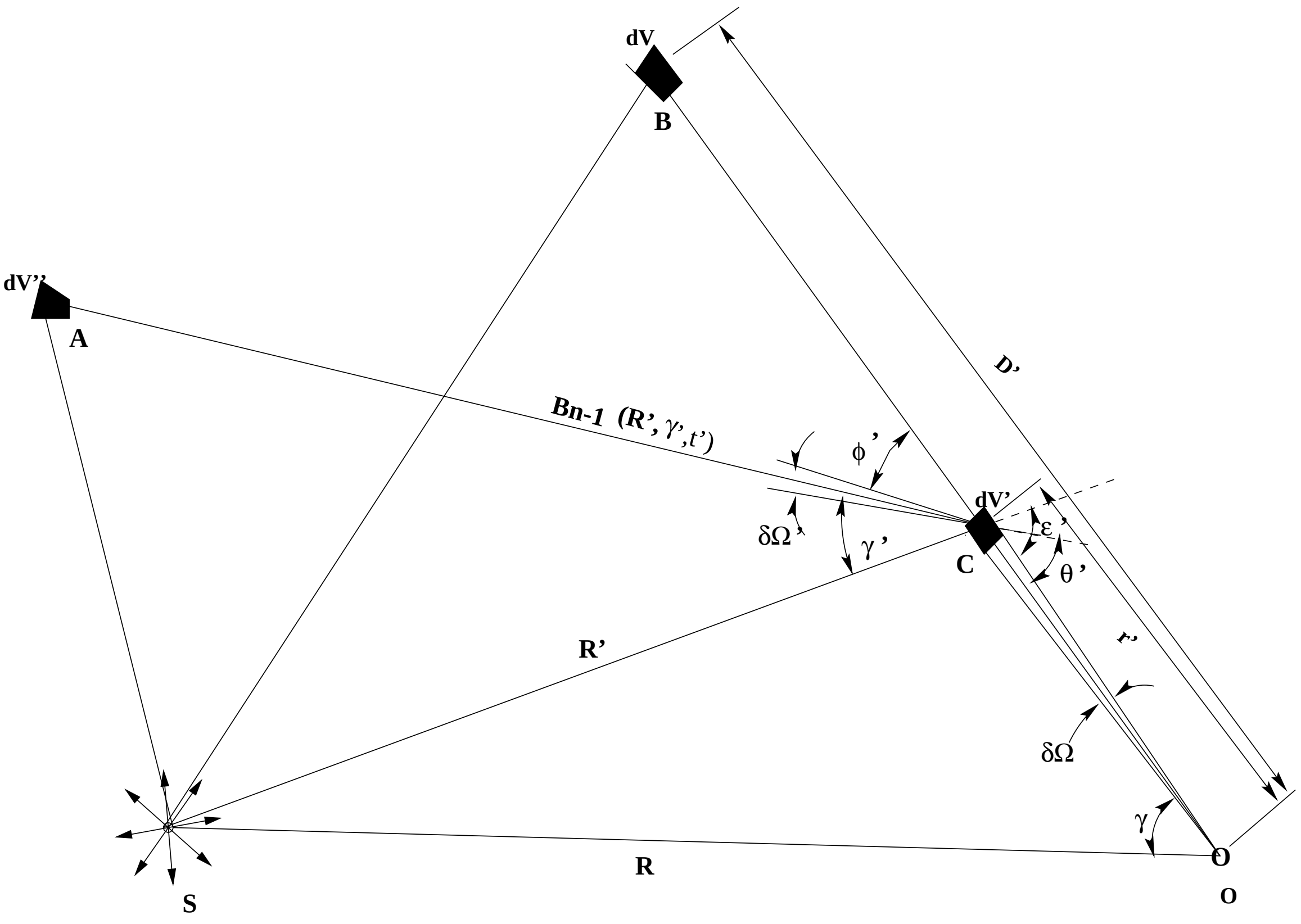

Figure 2: Geometric parameters involved in the calculation of $n$th-order scattering based on *(n-1)*th order scattering ( SOC and SOB are in the same plane; $\phi'$ is the the angle from the plane SAC to the plane SBC; $\theta'$ is the angle from AC to CO; $\epsilon'$ is the angle from SC to CO ).

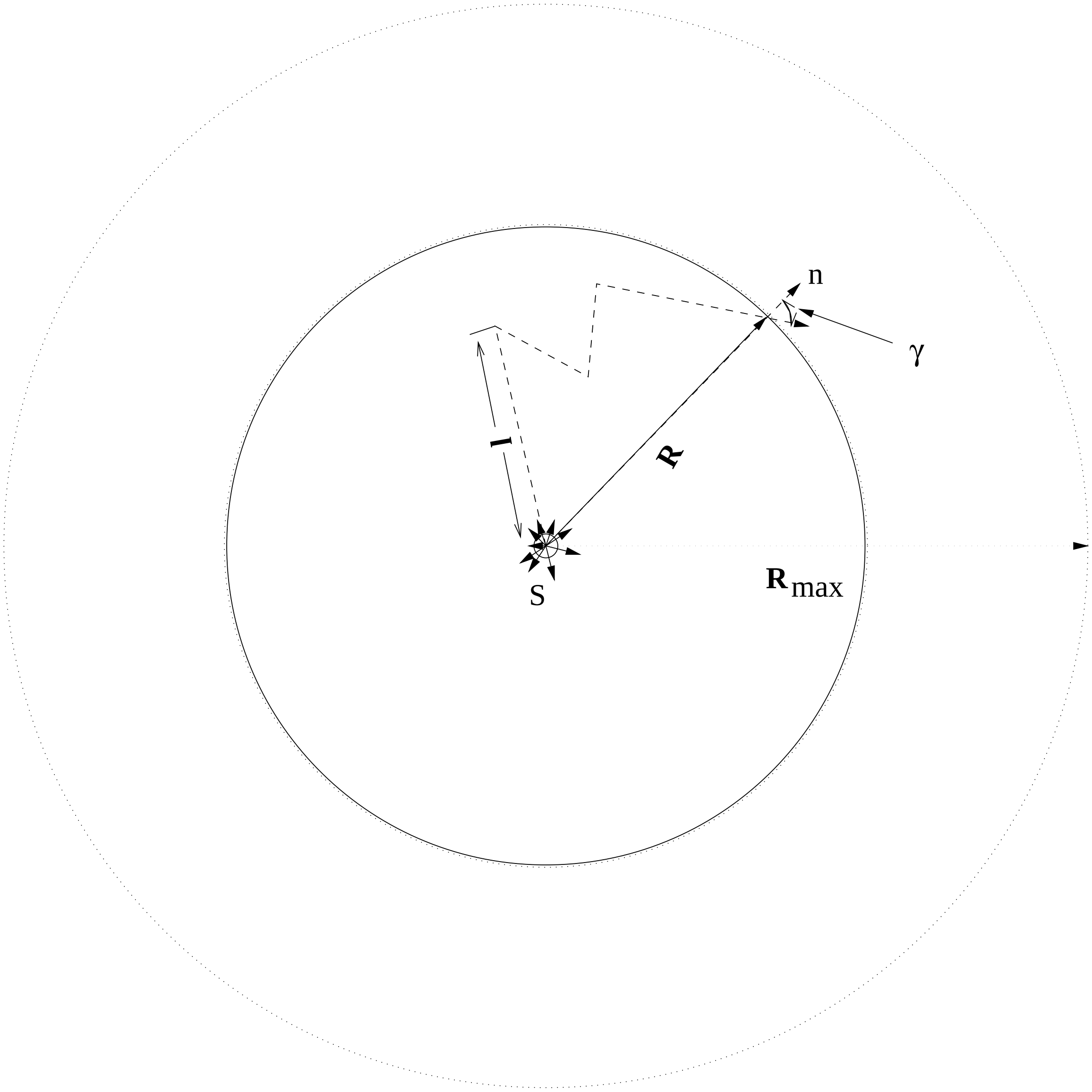

Figure 3: Geometric parameters involved in the MC simulation. $S$ is the source point. The sphere with radius R the detector. $n$ is the local norm direction of the detector.

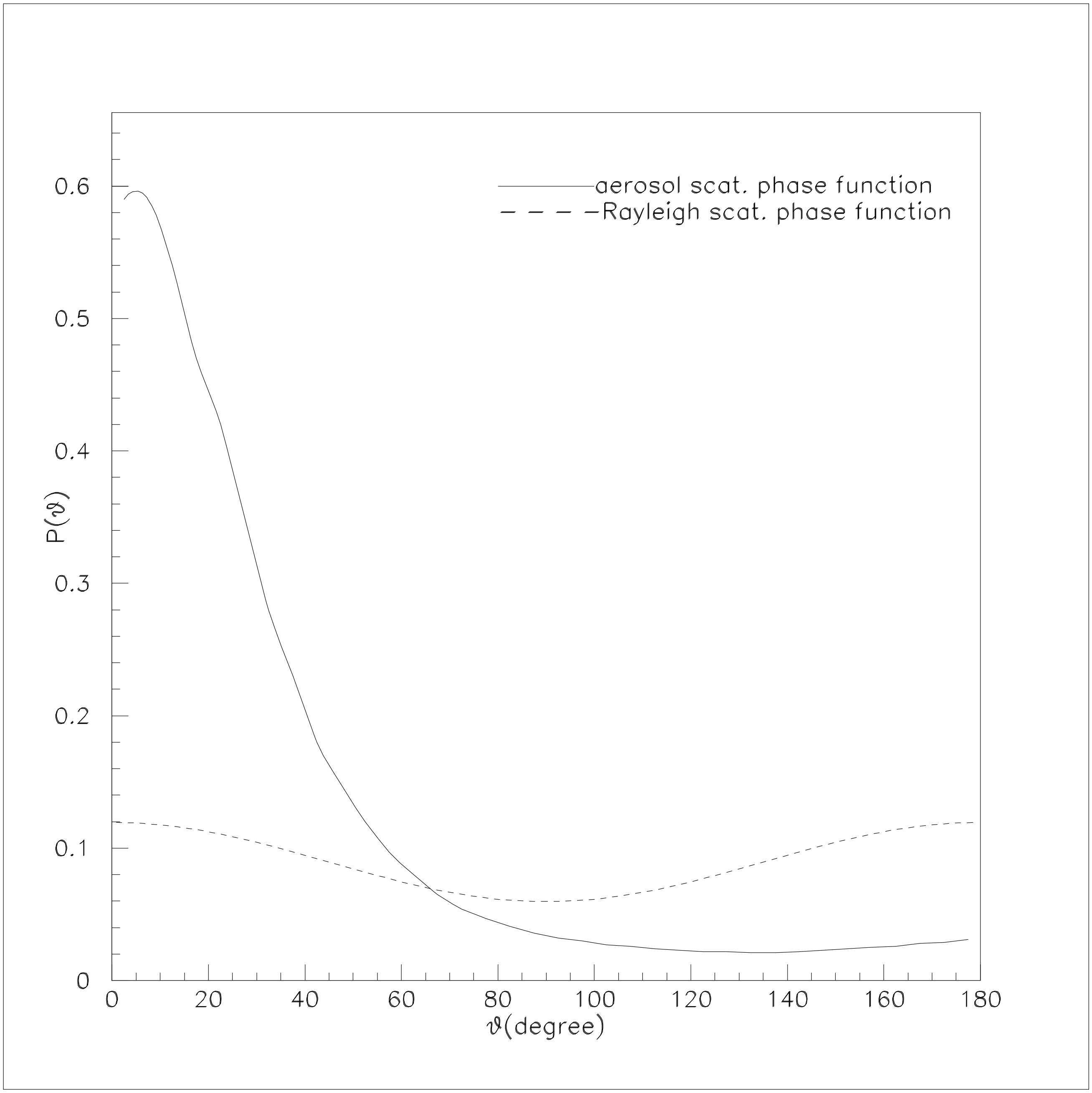

Figure 4: The normalized aerosol phase function used in the calculation. $\theta$ is from $0^o$ to $180^o$, $P(\theta)$ is the phase function.

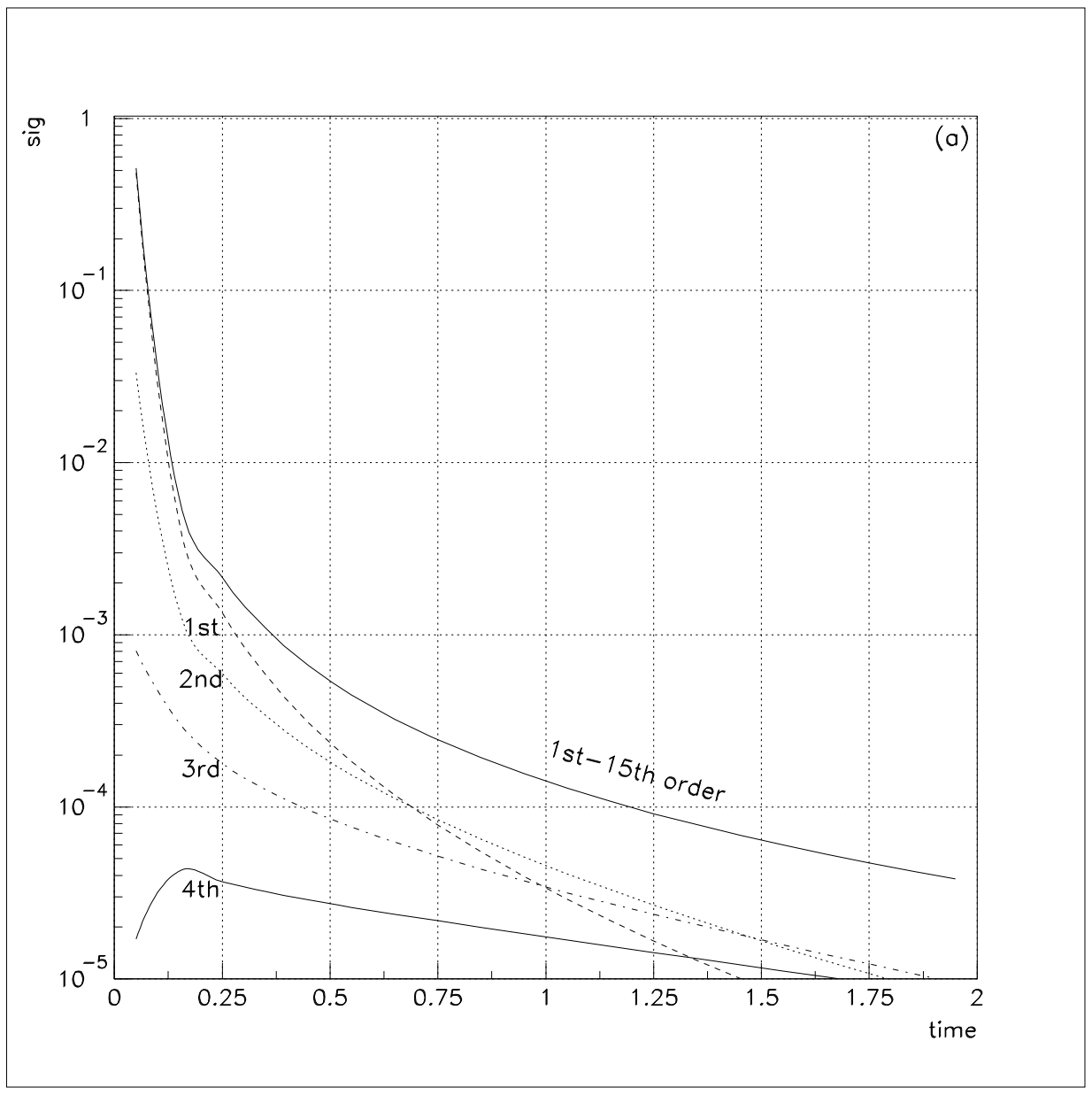

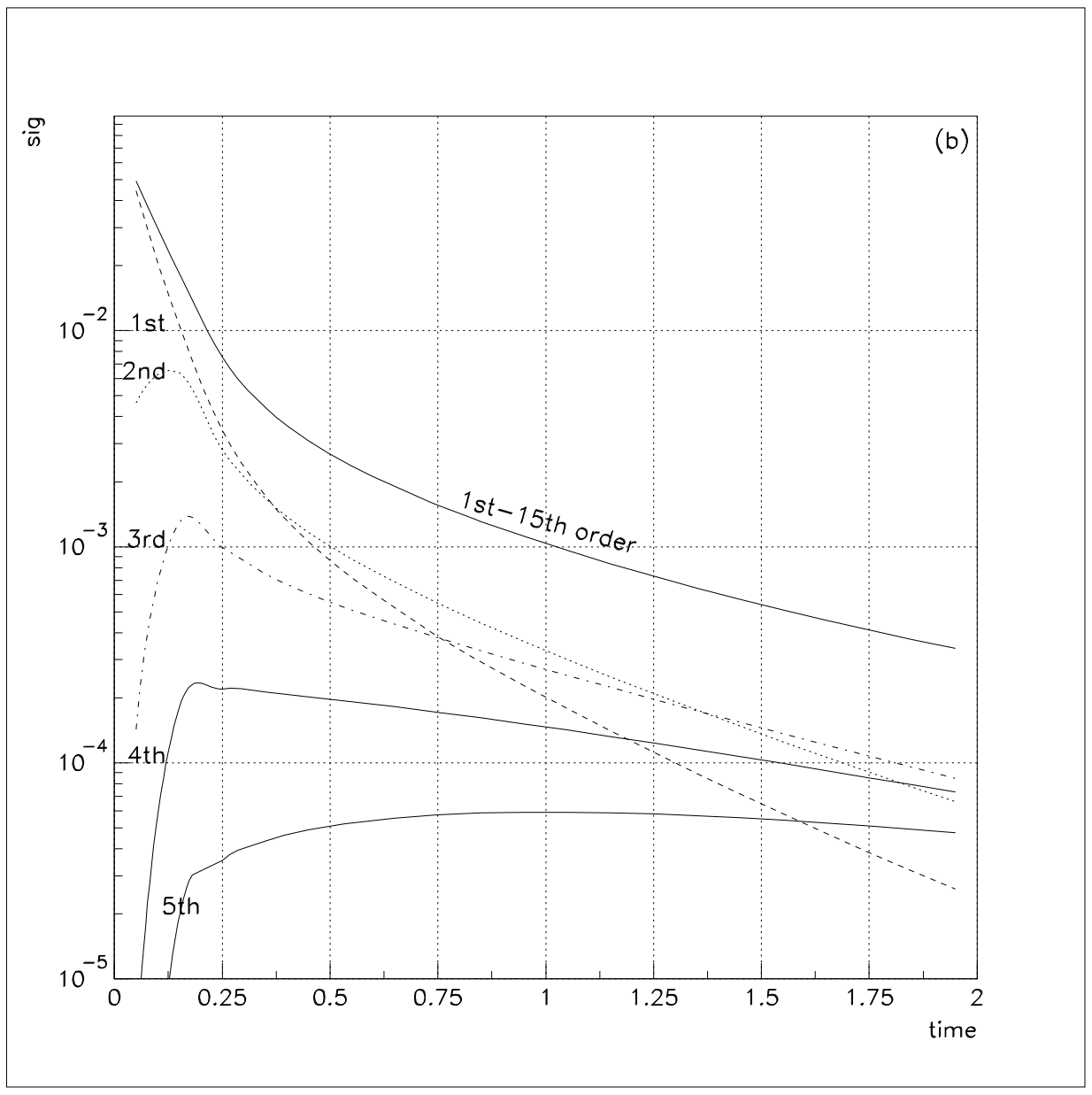

Figure 5: The first a few order of temporal scattering radiance and the sum of first 15th order of radiance. For $\rho = 0.8$,$\sigma = 0.8$,$R = 1$, $\gamma = 2^o$ for(a), and $\gamma = 20^o$ for (b). The X axis is the scattering light traveling time different from the direct light traveling time. The Y axis is the scattering signal

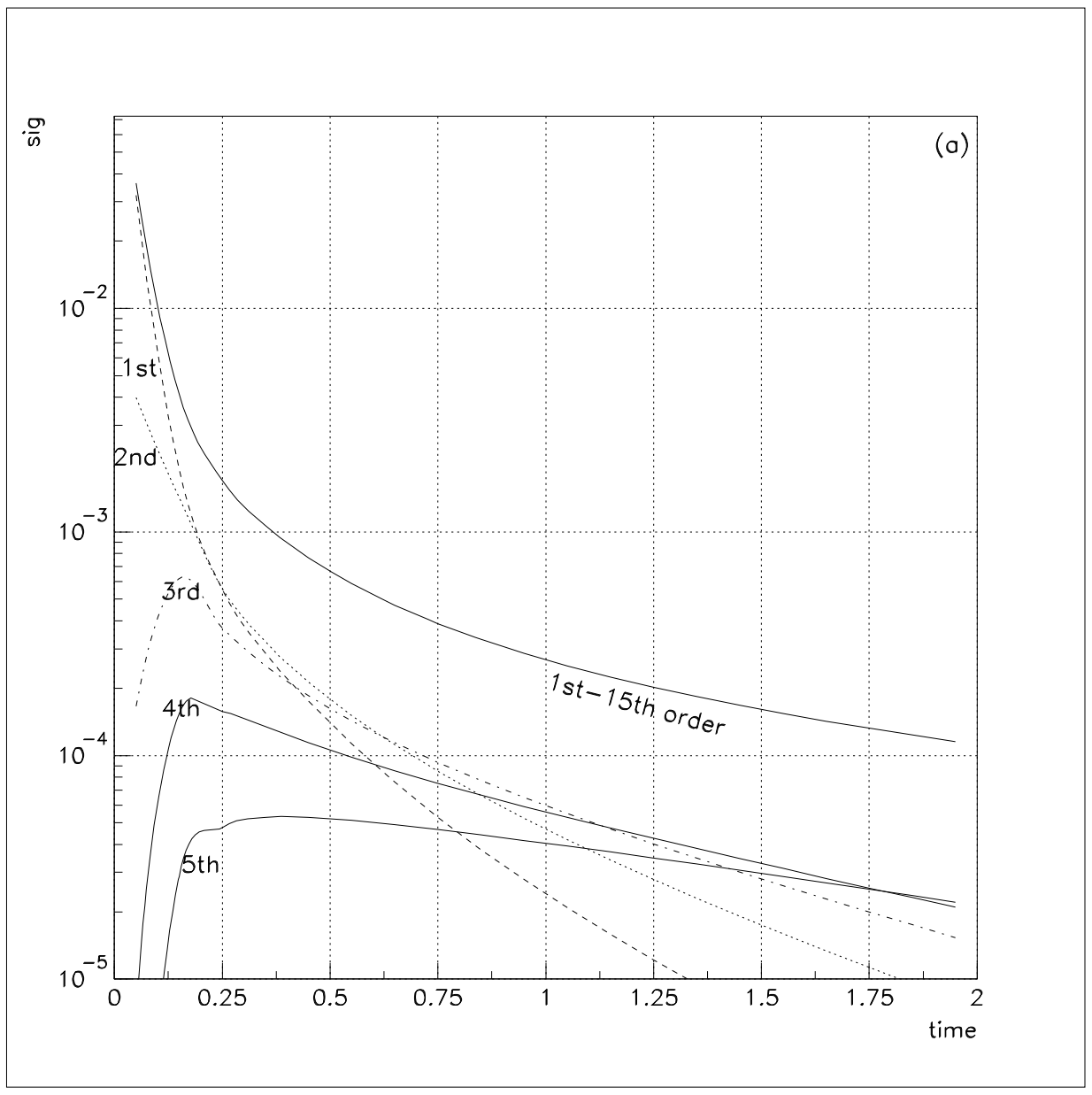

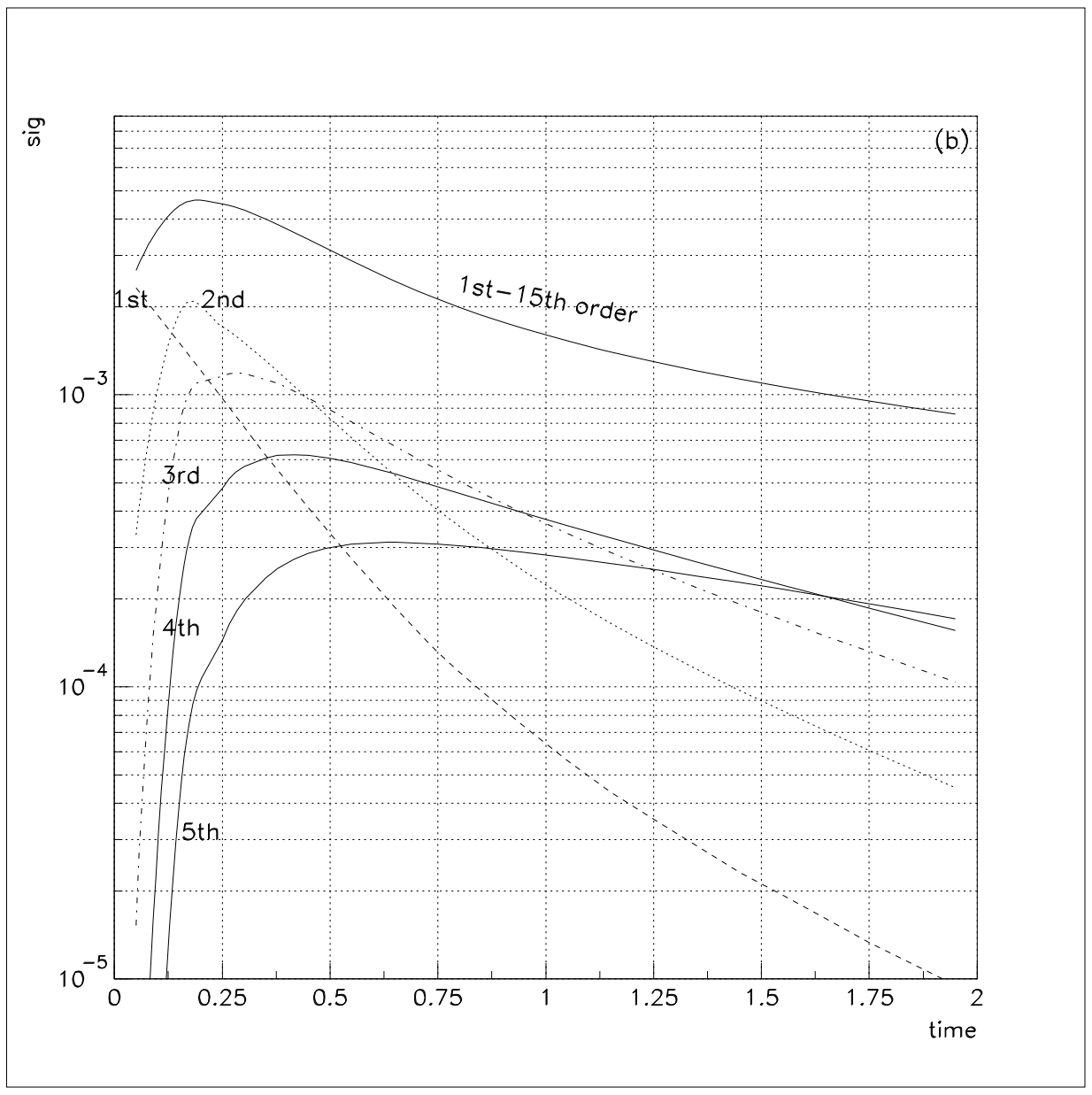

Figure 6: The first a few order of temporal scattering radiance and the sum of first 15 order of radiance. For $\rho = 0.8$,$\sigma = 0.8$,$R = 4$, $\gamma = 2^o$ for(a), and $\gamma = 20^o$ for (b).

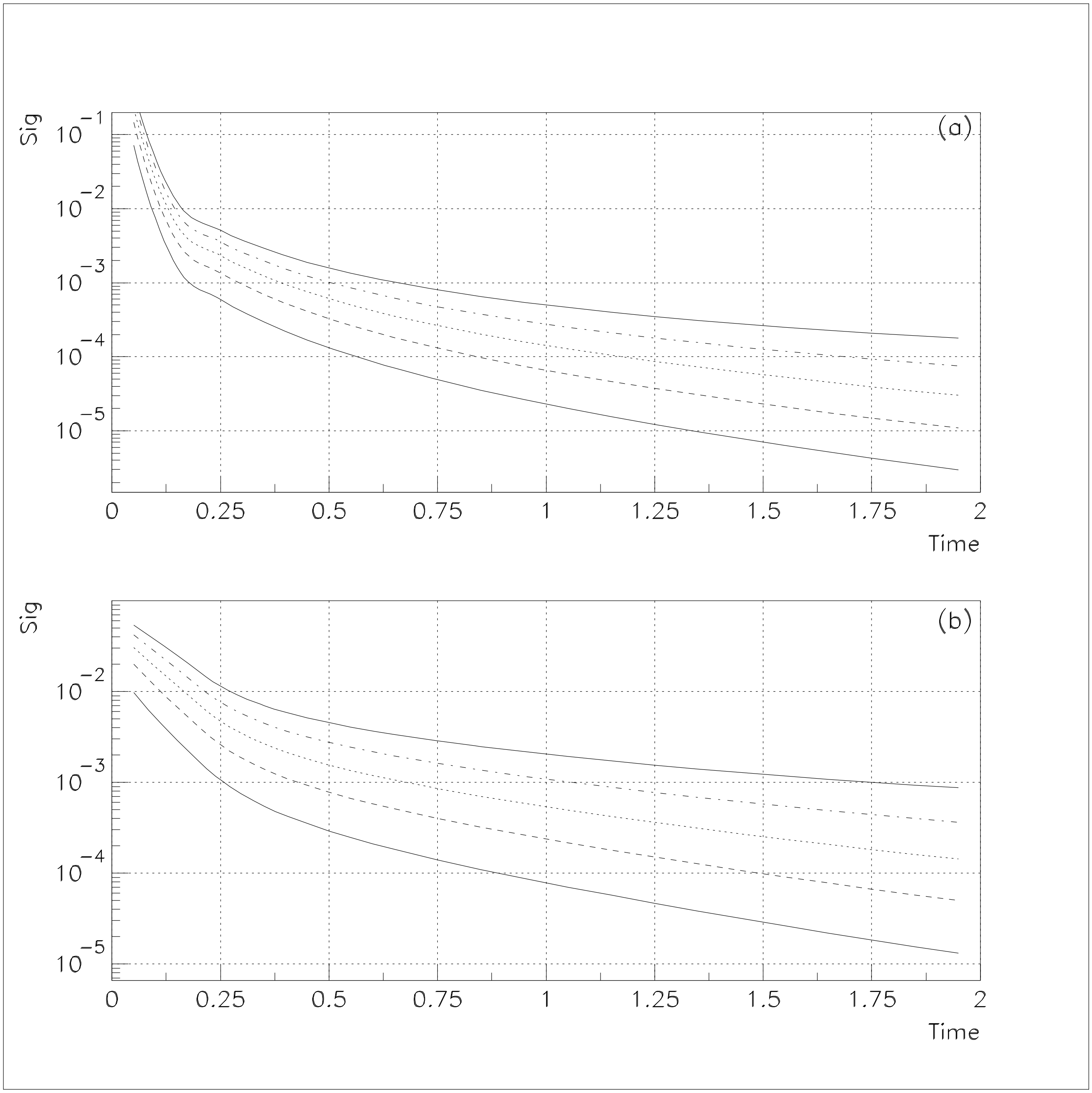

Figure 7: The dependency of the total scattering radiance on the single scattering albedo $\sigma$. $\sigma$ is from 0.2 to 1.0 from bottom to top. $R = 1$, $\rho = 0.8$, $\gamma = 2^o$ for(a), and $\gamma = 20^o$ for (b).

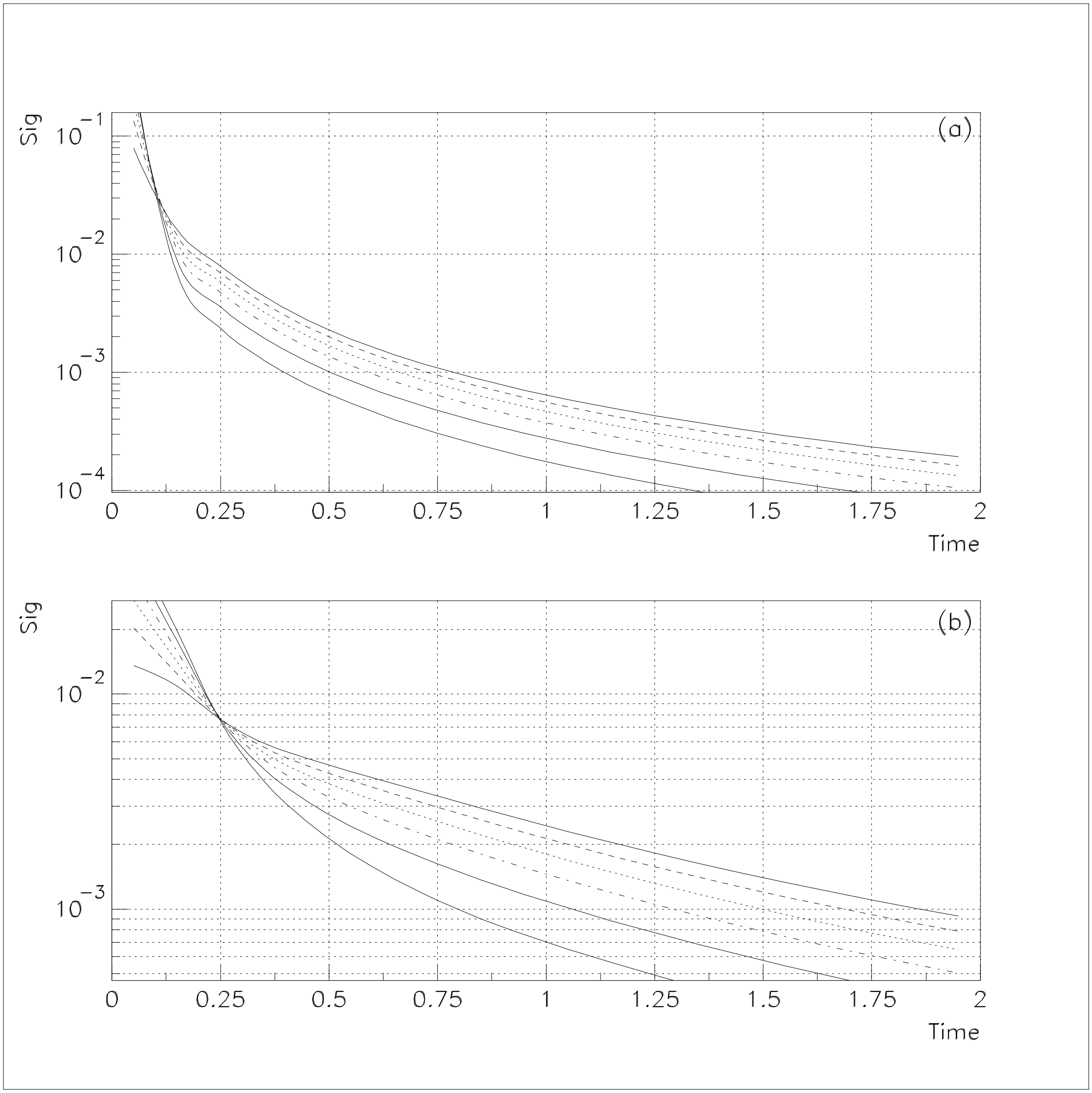

Figure 8: The dependency of the total scattering radiance on the rate of aerosol scattering coefficient $\rho$. $\rho$ is from 0.0 to 1.0 from top to bottom. $R = 1$, $\sigma = 0.8$, $\gamma = 2^o$ for(a), and $\gamma = 20^o$ for (b).

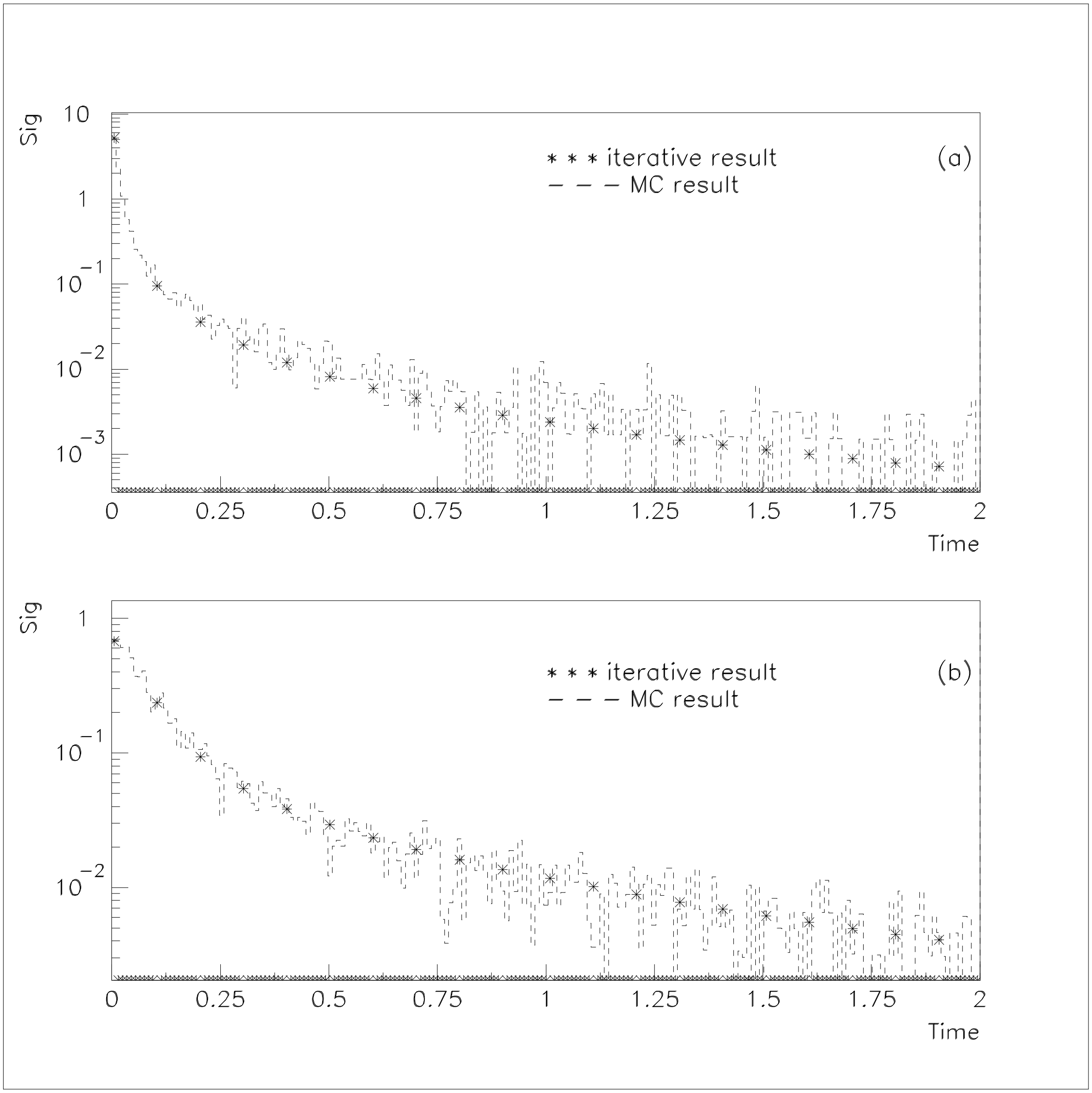

Figure 9: The comparisons of result from Monte Carlo calculations and the iterative method. $\rho = 0.8$, $\sigma = 0.8$, $R$ = *1*, $\gamma = 3^o$ for(a), and $\gamma = 19^o$ for (b).

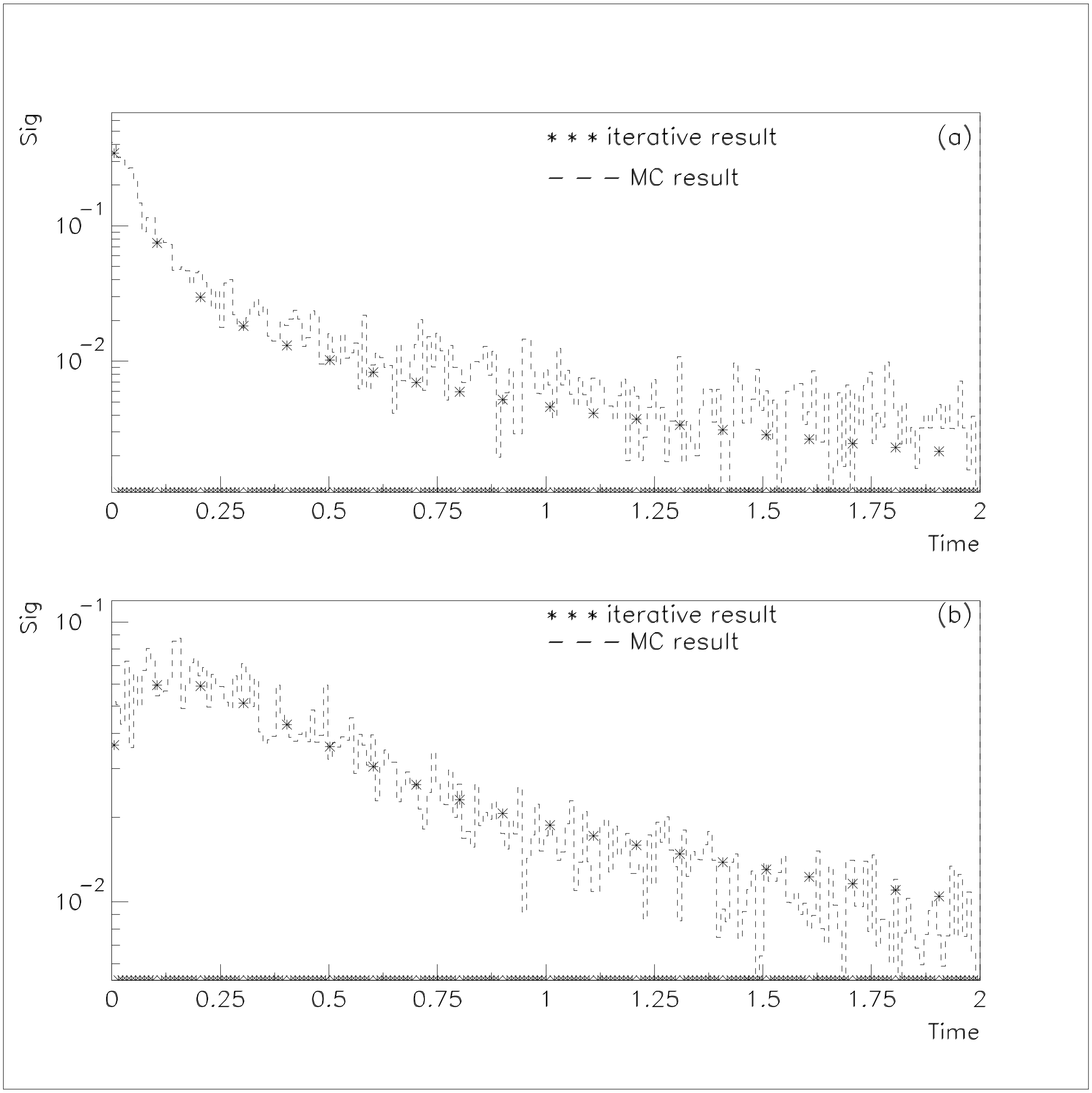

Figure 10: The comparisons of result from Monte Carlo calculations and the iterative method. $\rho = 0.8$, $\sigma = 0.8$, $R = 4$, $\gamma = 3^o$ for(a), and $\gamma = 19^o$ for (b).

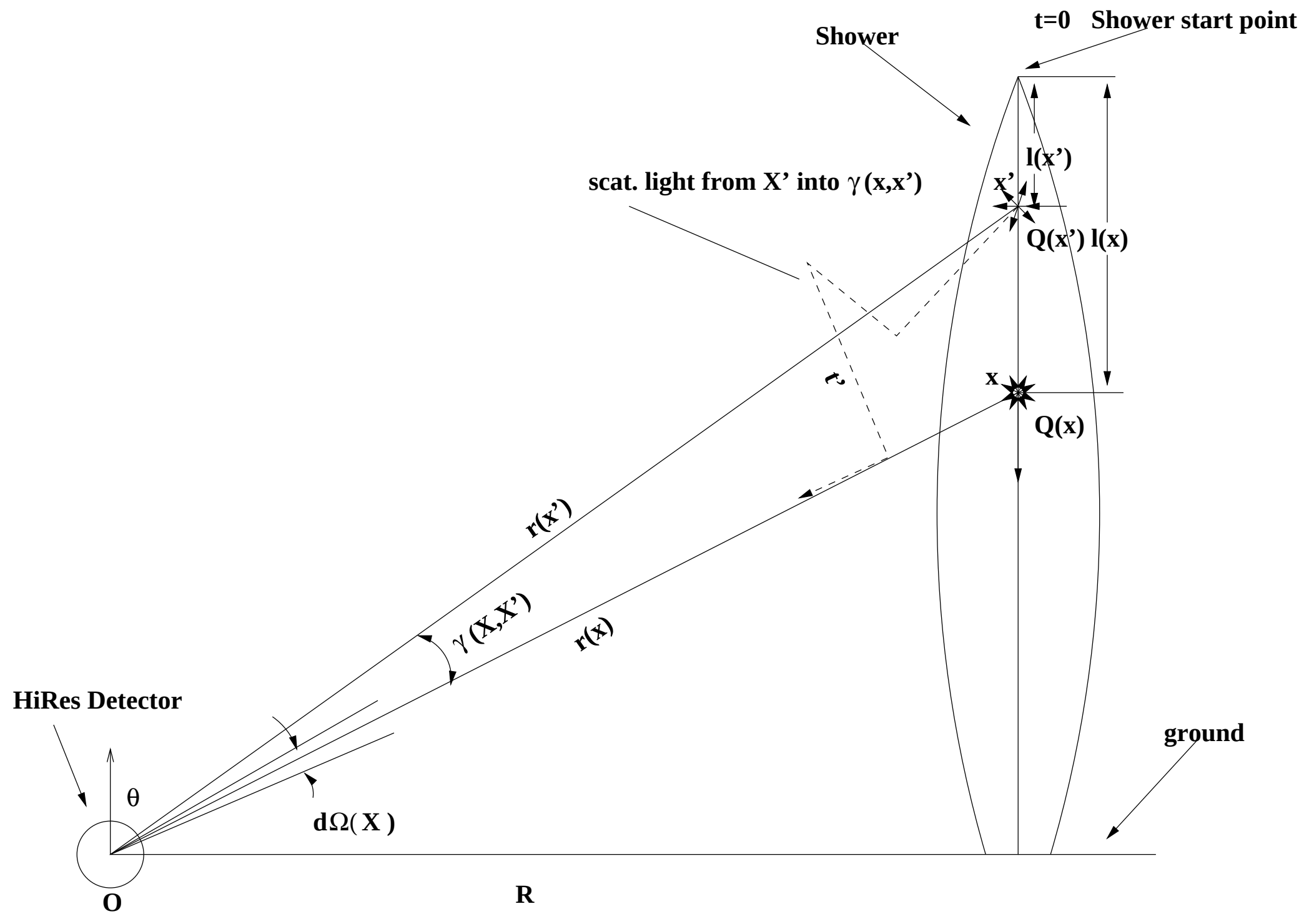

Figure 11: The Geometric parameters involved in the calculation of multiply scattering effect for an EAS. $X'$ is the source point, with total number of photons $\mathcal{Q}(x')dx'$. The time clock is set to 0 at the start point of shower, $t'$ is the real time the photon from $X'$ point hit the detector with $\gamma(x', x)$ angle off from source $X'$.

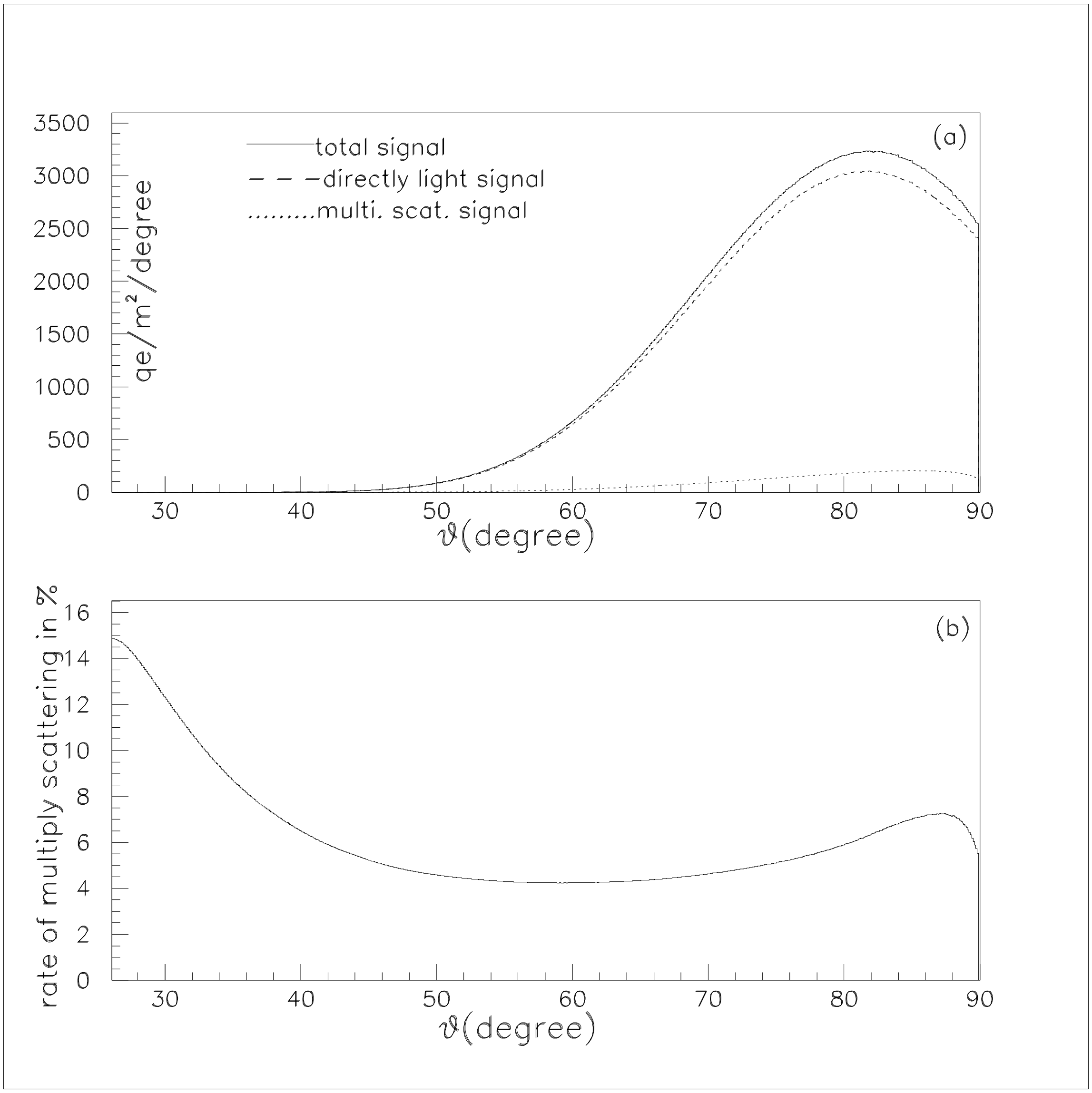

Figure 12: The effect of multiply scattering on vertical the EAS shower 10 Km away from the detector. (a) The EAS shower longitudinal profile seen from HiRes detector. X axis is the zenith angle in degree; Y axis is the signal detected from HiRes site. Because of the threshold of the detector, only the bins $\pm 10^o$ from the maximum point will trig the detector. (b) The rate of multiply scattering signal to the directly light signal.

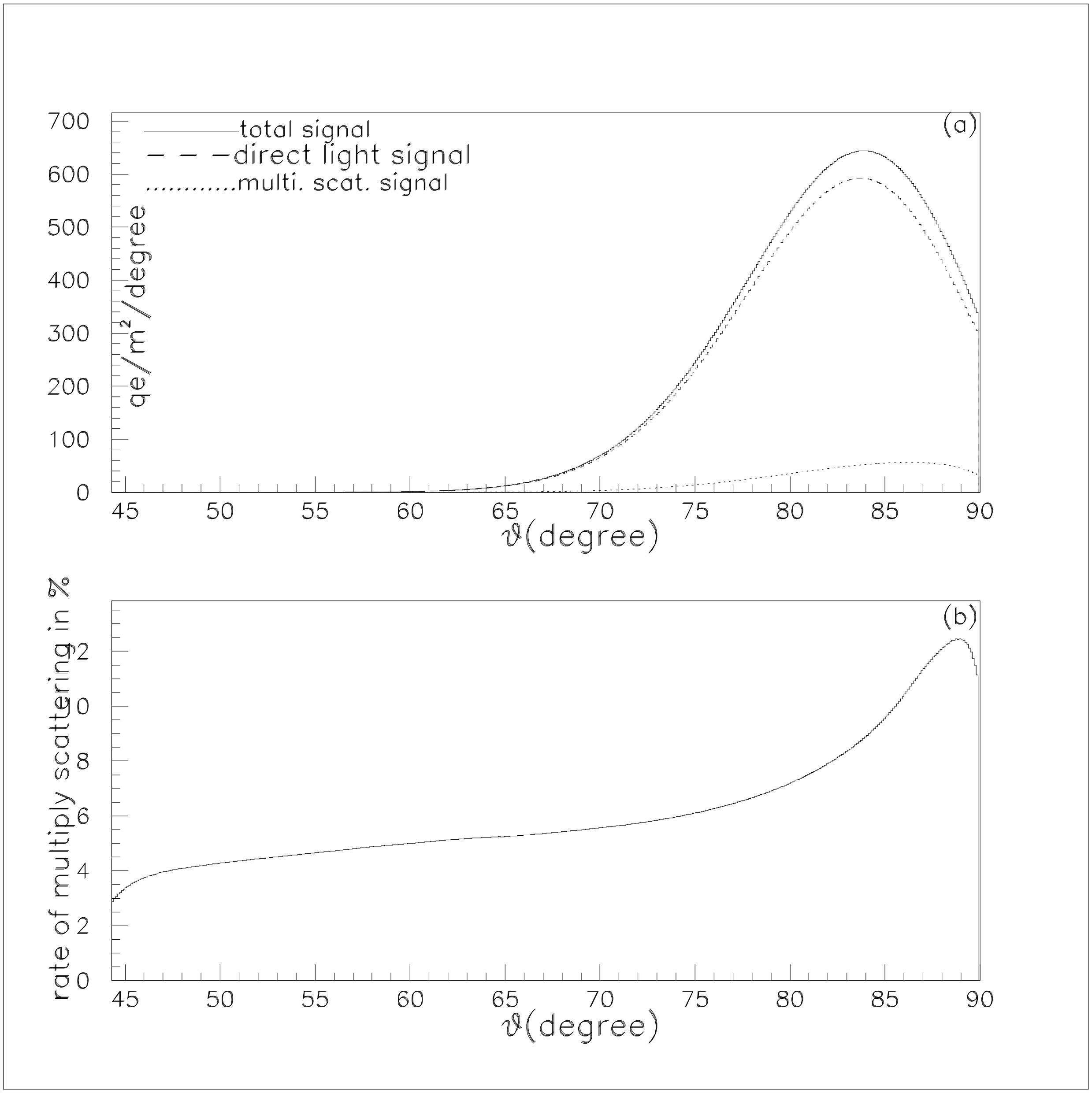

Figure 13: Same as Fig. 12. except the shower to detector distance is 20 Km

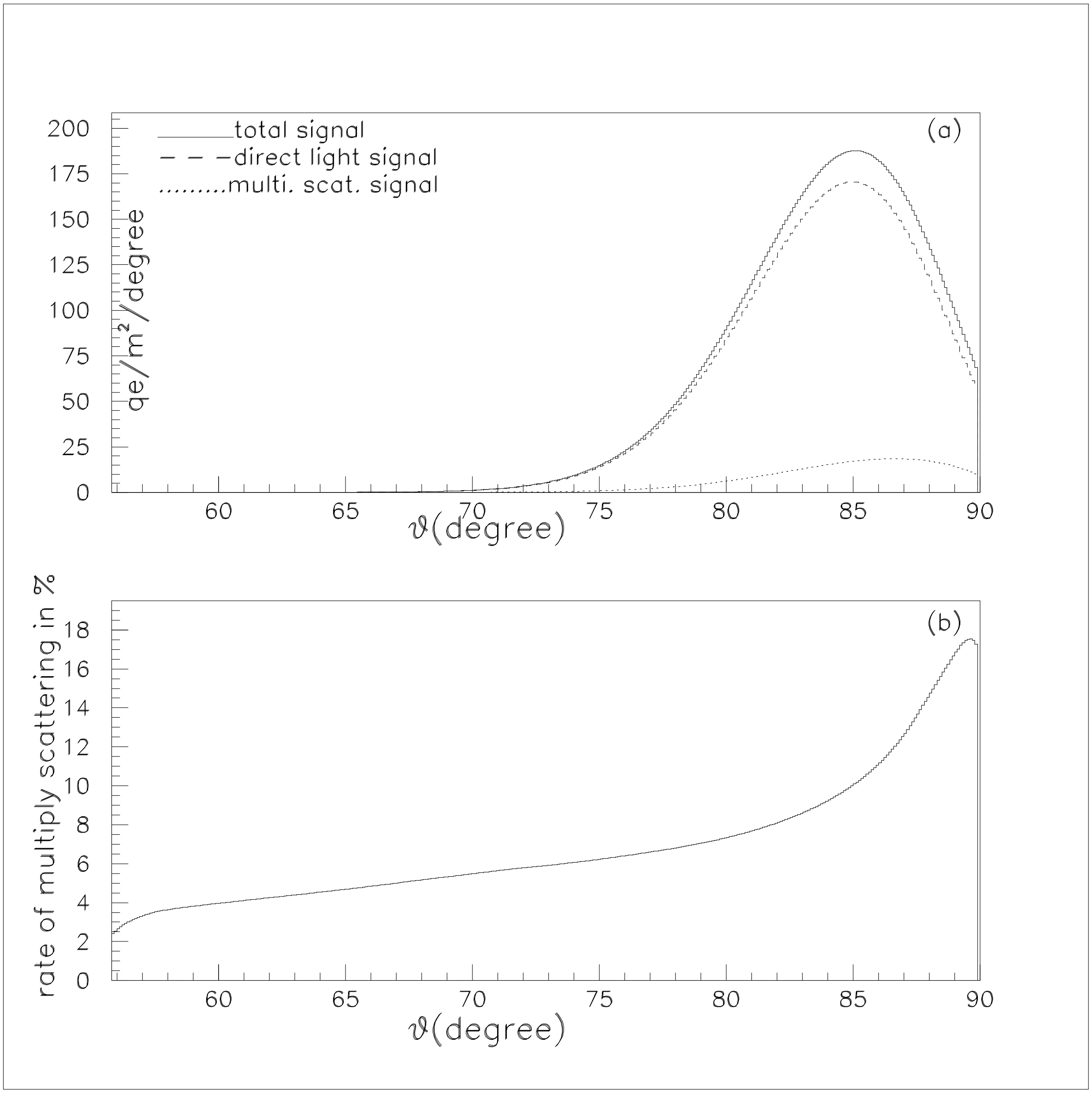

Figure 14: Same as Fig. 12. except the shower to detector distance is 30 Km

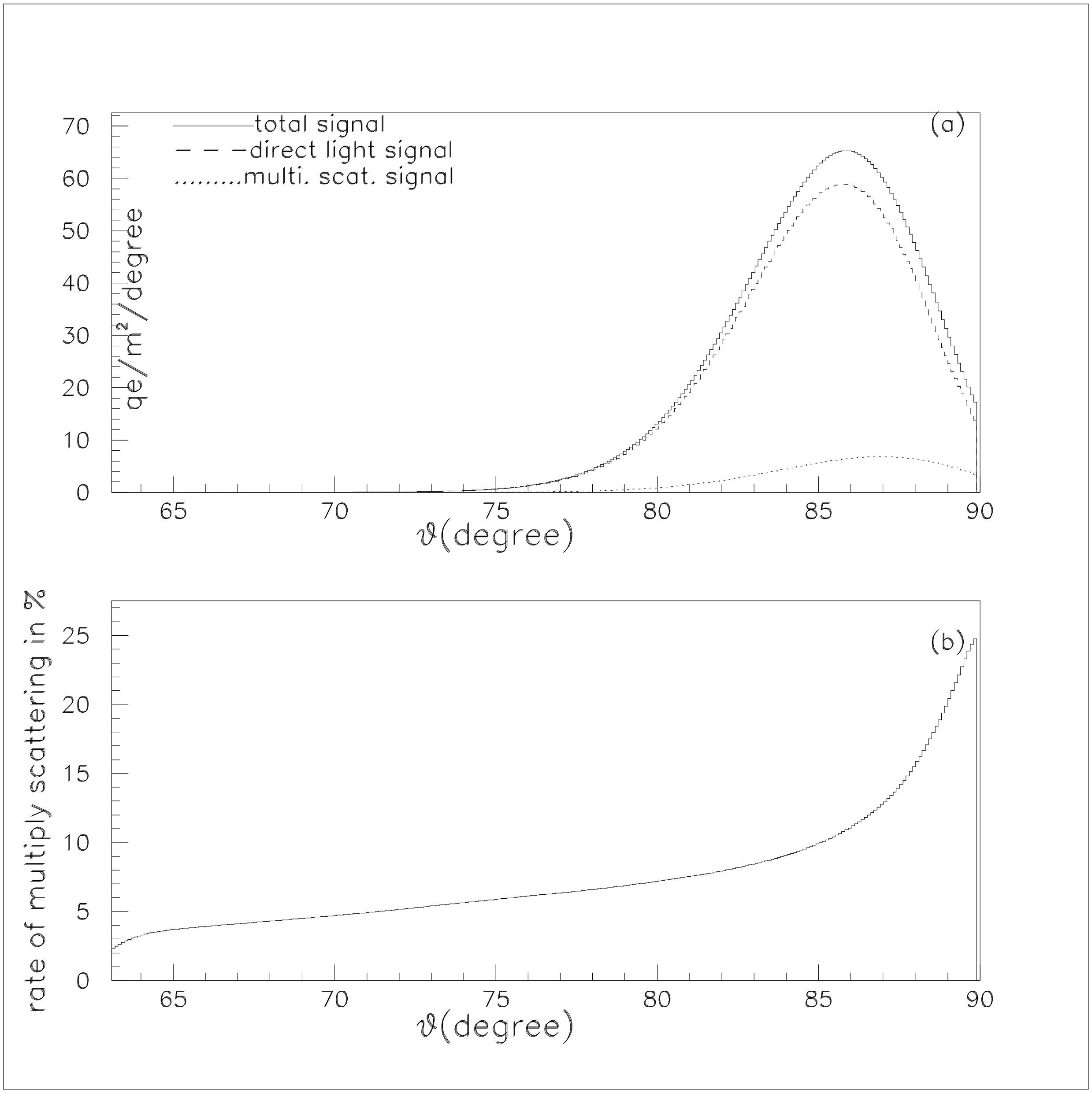

Figure 15: Same as Fig. 12. except the shower to detector distance is 40 Km

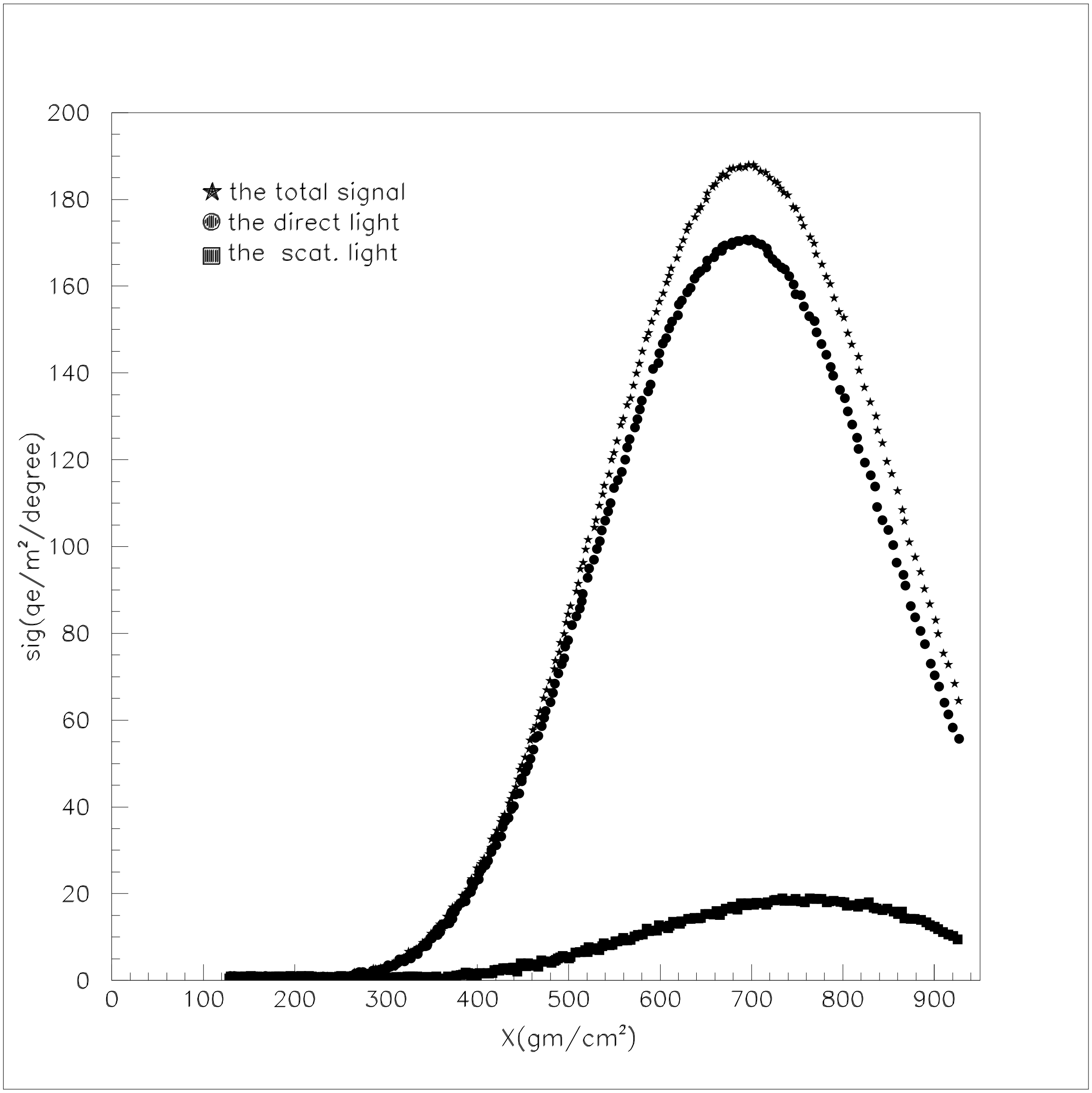

Figure 16: The same shower longitudinal profile as in Fig. 14. The X axis is the slant depth of the shower at every stage. The total aerie under the line proportional to the primary energy of the Cosmic ray.

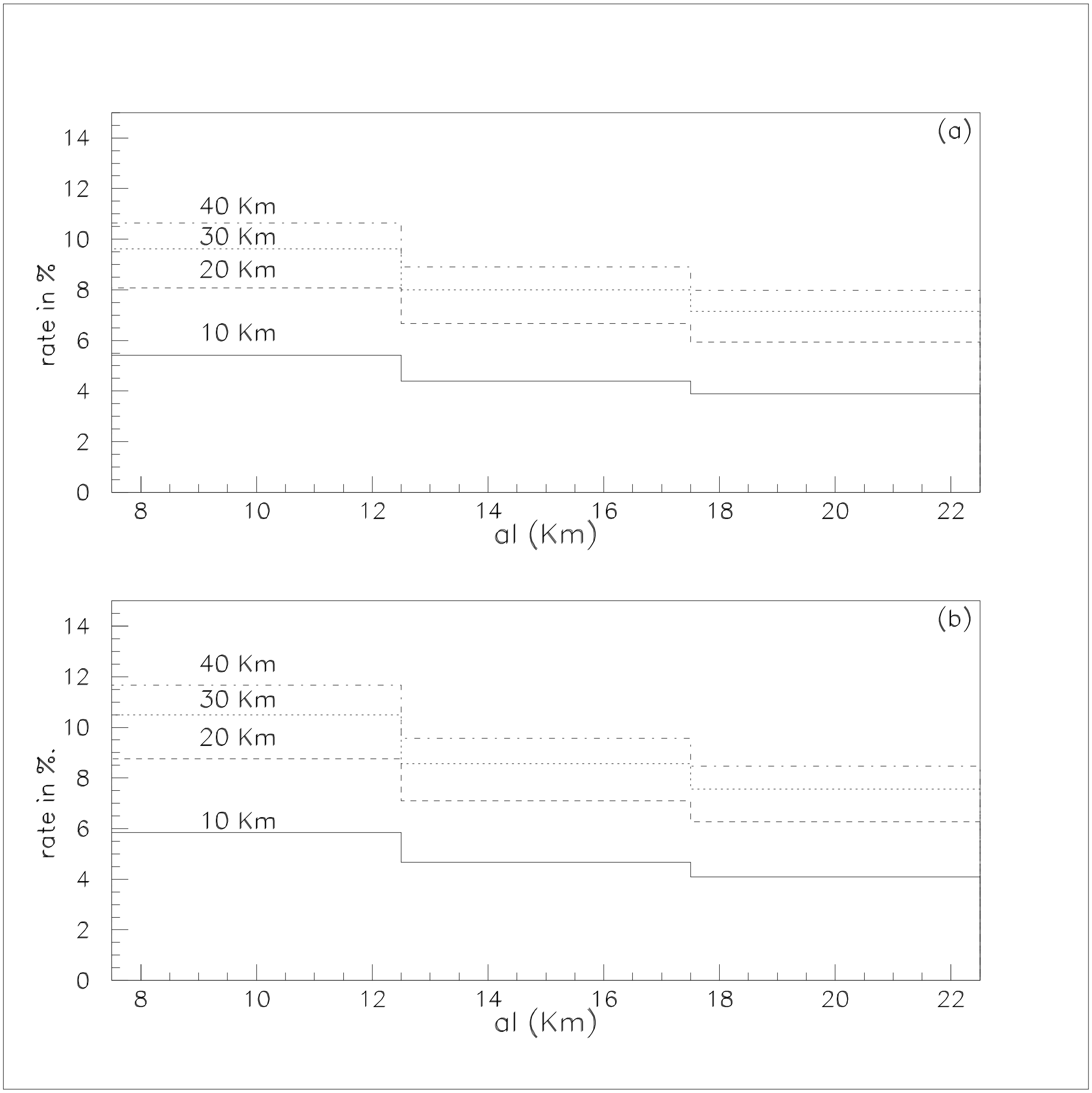

Figure 17: The contribution of the multiply scattering light to the primary energy estimation in percentage depends on different atmosphere model for different shower detector distance. X axis is the aerosol horizontal extinction length. (a) the aerosol scale height is 1.2 Km. (b)the aerosol scale height is 1.5 Km.